\documentclass{ifacconf}

\def\originalspace{1} %comment for short space
\def\longfigure{1} %comment for short figure data dictionary 

\usepackage{xspace}
\usepackage{graphicx}      % include this line if your document contains figures
\usepackage{natbib}        % required for bibliography
\usepackage{blindtext}
\usepackage[normalem]{ulem}
\usepackage{amsmath, amssymb, amsfonts}
\usepackage{mathrsfs}
\usepackage{dsfont}
\usepackage{xcolor}
\usepackage{mathrsfs}
\usepackage{tikz}
\usepackage{tikz-cd}
\usepackage{mathtools}
\ifx\originalspace\undefined%
\newcommand{\shortsection}[1]{\vspace{-1mm}\section{#1}\vspace{-2.5mm}}

\newcommand{\shortsubsection}[1]{\vspace{-1mm}\subsection{#1}\vspace{-2mm}}
\else
\newcommand{\shortsection}[1]{\section{#1}}
\newcommand{\shortsubsection}[1]{\subsection{#1}}

\fi

\ifx\longfigure\undefined%
\newcommand{\figurenameacademic}{acadm_f1_sh}
\newcommand{\figurenameunbalanc}{unbal_f1_sh}
\else
\newcommand{\figurenameacademic}{acadm_f1_lg}
\newcommand{\figurenameunbalanc}{unbal_f1_lg}
\fi

\newcommand{\tss}[1]{\textsuperscript{#1}}

\newcommand{\mc}[1]{\mathcal{#1}}
\newcommand{\mf}[1]{\mathfrak{#1}}
\newcommand{\mr}[1]{\mathrm{#1}}
\newcommand{\mb}[1]{\mathbb{#1}}

\newcommand{\msf}[1]{\mathsf{#1}}
\newcommand{\mt}[1]{\mathtt{#1}}

\newcommand{\col}{\mr{col}}
\newcommand{\diag}{\mr{diag}}
\newcommand{\rank}[1]{\mr{rank}\left(#1\right)}

\newcommand{\coldef}[1]{\mathrm{col}\!\left(#1\right)}

\newcommand{\unaryminus}{\scalebox{0.65}[1.0]{\ensuremath{\,-\,}}}

\newcommand{\kron}{\otimes} % just rename
\newcommand{\bkron}{\circledcirc}

\newcommand{\datasetk}[2]{\{#2\}_{k=1}^{#1}}

\newcommand{\svdots}{\raisebox{0pt}{$\scalebox{.75}{\vdots}$}}
\newcommand{\sddots}{\raisebox{0pt}{$\scalebox{.75}{$\ddots$}$}}

\newcounter{ass}

\newcommand{\comment}[1]{}

\begin{document}
\begin{frontmatter}

\title{Data-Driven Predictive Control for Linear Parameter-Varying Systems\thanksref{footnoteinfo}} 
% Title, preferably not more than 10 words.

\thanks[footnoteinfo]{This work has received funding from the European Research Council (ERC) under the European Union's Horizon 2020 research and innovation programme (grant agreement nr. 714663), the European Space Agency in the scope of the `AI4GNC' project with SENER Aeroespacial S.A. (contract nr. 4000133595/20/NL/CRS), and the Deutsche Forschungsgemeinschaft (DFG, German Research Foundation)-Project No. 419290163. Corresponding author: Chris Verhoek (\texttt{c.verhoek@tue.nl})}

\author[tue]{Chris Verhoek} 
\author[ieem]{Hossam S. Abbas} 
\author[tue,sztaki]{Roland T\'oth}
\author[tue]{Sofie Haesaert} 

\address[tue]{Control Systems Group, Dept. of Electrical Engineering, Eindhoven University of Technology, Eindhoven 5600MB, The Netherlands}
\address[ieem]{Institute for Electrical Engineering in Medicine, Universit{\"a}t zu L{\"u}beck, 23558 L{\"u}beck, Germany}
\address[sztaki]{Systems and Control Lab, Institute for Computer Science and Control (SZTAKI), Kende u. 13-17, 1111 Budapest, Hungary}

\begin{abstract}                % Abstract of not more than 250 words.
Based on the extension of the behavioral theory and the Fundamental Lemma for Linear Parameter-Varying (LPV) systems, this paper introduces a Data-driven Predictive Control (DPC) scheme capable to ensure reference tracking  and satisfaction of Input-Output (IO) constraints for an unknown system under the conditions that (i) the system can be represented in an LPV form and (ii) an informative data-set containing measured IO and scheduling trajectories of the system is available. It is shown that if the data set satisfies a persistence of excitation condition, then a data-driven LPV predictor of future trajectories of the system can be constructed from the IO data set and online measured data. The approach represents the first step towards a DPC solution for nonlinear and time-varying systems due to the potential of the LPV framework to represent them. Two illustrative examples, including reference tracking control of a nonlinear system, are provided to demonstrate that the data-based LPV-DPC scheme, achieves similar performance as LPV model-based predictive control.

\ifx\originalspace\undefined \vspace{-3mm} \fi
\end{abstract}

\begin{keyword}
Predictive Control; Data-Driven Control; Linear Parameter-Varying Systems; Non-Parametric Methods.
\end{keyword}

\end{frontmatter}

%===============================================================================
\shortsection{Introduction}
Due to the increasing complexity of systems in engineering, designing control solutions based on traditional modeling methods is becoming more and more challenging. Deriving models based on first principle laws for complex systems is costly and cumbersome, and often held back by unknown dynamic details and the difficulty to decide which physical phenomena are important to address for the control task at hand. Therefore, in the past decades, various approaches have been researched to either simplify or automate the modeling and control of these complex systems. The aspect of learning has often been used to accommodate the simplification or the automation steps. %However, most classical contributions to modern control theory rely on an analytic, mathematical description of the system to be controlled, such that we can design a controller which ensures stability and performance properties for the closed-loop system. 

Learning complex systems from data was introduced in the systems and control community in terms of system identification with the objective to recover the relevant dynamic relations of the system directly from data. Based on the estimated models, controllers can be designed to guarantee stability and performance properties for the \emph{model-based} closed-loop system. However, how these guarantees apply on the \emph{actual} system largely depends on estimation error, i.e. modeling uncertainty, introduced by the identification and represents a challenging problem that has been in the focus of intensive research \cite{Ljung1999,oomen2013connecting,zhou1996robust}.

%%%%%%%%%%%%%%%%%%%%%%%%%%%%%%%%%%%%%%%%%%%%%%%
%{\color{red} xxxxx}
%
%{\color{red} Just continue here with that it is attractive to avoid this difficulty and instead of the separate modelling and controller design step directly synthesise or estimate a controller from data. Mention dual control?}
%
%
%{\color{red} What you write here is problematic as one part of robust control claims that they do solve this problem, and you don't even recongise that....}
%There are some recent advances that can give guarantees based on e.g. the modeling error or uncertainty sets (see e.g. \cite{dean2019sample}), but a characterization of how the modeling error affects the model based design remains unknown. Furthermore, there are many practical applications where identifying a model of the physical system is difficult, costly and/or time-consuming. Hence, an important question within the control community is how to design a stabilizing and performing controller directly from data. Recently, this question has gained a lot of attention in the works on data-driven control, see e.g. \cite{HouWang2013} for a survey.

Data-driven control can bypass one of the most tedious steps of model-based controller design approaches, which is to obtain an appropriate model of the plant to be controlled. Data-driven control approaches directly determine control laws and policies from data. This idea resulted in many contributions over the years for many classes of systems \cite[]{HouWang2013}, including Linear Parameter-Varying (LPV) systems \cite[]{formentin2013direct,Formentin2016direct}. However, guarantees for stability and performance of the closed-loop are lacking for most of these methods. Furthermore, how predictive control schemes and data-driven representations are integrated in the control framework are largely unknown for systems beyond the Linear Time-Invariant (LTI) class.

Most of the aforementioned problems are tackled for LTI systems in recent years \cite{RomerBerberichKohlerAllgower2019, CoulsonLygerosDorfler2019, dePersisTesi2019journal}. In the context of behavioral systems theory, \cite{WillemsRapisardaMarkovskyMoor2005} characterize the full behavior of a data-generating LTI system (when focusing on trajectories of a certain length) purely based on measured Input-Output (IO) data and the condition that the input is persistently exciting. Based on this persistently excitation condition of the data, a new generation of algorithms for data-driven control have been developed, e.g. \cite{MaRa08}. This principle has been successfully fused with data-driven predictive control in \cite{CoulsonLygerosDorfler2019}. Furthermore, the problem of data-driven simulation, i.e., simulation based on only input-output data, has been extended in \cite{BerberichAllgower2020} into certain classes of nonlinear systems where linearity plays an important role, i.e. special cases of Hammerstein and Wiener systems. Moreover, a method for verifying dissipativity properties of a system from input-output data-trajectories has been introduced in \cite{RomerBerberichKohlerAllgower2019}, which can be a basis for establishing stability and performance guarantees in such data-driven framework. In this regard, \cite{dePersisTesi2019journal, DePersisTesi2019} have studied related connections to the classic Lyapunov stability, which allows Linear Matrix Inequalities for designing state-feedback controllers. Considering noisy data has also been addressed in several of the aforementioned references. However, the predominant feature of most of these approaches is that they deal with LTI systems, which may be merely an approximation of the increasingly complex system behaviors in practice.

The LPV systems framework \cite[]{To10} has the potential of describing nonlinear and time-varying behaviors using a \emph{linear} dynamic structure, which depends on a so-called scheduling variable associated with some measurable exogenous, or endogenous signals of the system. The scheduling signal affects the operating point of the system and can be used to schedule online controllers designed for the system based on linear optimal approaches. Therefore, LPV controllers have received considerable attention, see e.g., \cite{HoWe15}. In this paper, we aim to take the first steps towards a nonlinear data-driven system formulation by considering LPV systems framework.

Predictive control \cite{Ma02j} is an online control approach that can systematically handle systems constraints. Its paradigm is to solve an online optimization problem to optimize the system performance  based on a short term prediction of its future behavior. Therefore, linear formulation of the system dynamics is preferred to avoid computational complexity and hence LPV model-based predictive control (LPV-MPC) has become an attractive procedure for controlling nonlinear and time-varying systems \cite[]{MoNoSe20}. However, an accurate LPV model for the system  should be available, which is usually difficult to obtain and affected by the same model uncertainty as in the LTI case. On the other hand, predictive control became a host of most learning-based control methodologies, e.g., \cite{HeWaMeZe20}, due to its appealing feature of online data mining and its capability to incorporate safety guarantees.

In this paper, our contribution is to propose a Data-driven Predictive Control (DPC) scheme for LPV systems capable to ensure reference tracking and satisfaction of IO constraints for an unknown LPV system. The method is based on the extension of the behavioral theory and the Fundamental Lemma for the LPV system class and allows to construct a data-driven LPV predictor of future trajectories from previously recorded data set of the system, satisfying a Persistence of Excitation (PE) condition. The approach represents the first step towards a DPC solution for nonlinear and time-varying systems, due to the capability of the LPV framework to represent them, and holds the potential to generalize the data-driven stability and performance guarantees of the LTI case.

The paper is structured as follows. In Section \ref{sec:preliminaries}, some preliminaries are introduced, while Section \ref{sec:analysis} establishes our main contribution in terms of the LPV data-driven predictor based on the Fundamental Lemma. Based on these results, our second contribution in terms of introducing an LPV-DPC scheme is discussed in Section \ref{sec:dpc}. The effectiveness of our approach is shown in Section \ref{sec:example} by means of simulation studies on an academic example and an unbalanced disc setup. The conclusions and outlooks are given in Section \ref{sec:conclusion}.

\textit{Notation:}
For a discrete-time signal $s:\mb{N}\to\mb{R}^{n_\mr{s}}$ with $n_\mr{s}>0$ and $\mb{N}$ the set of positive integers, we denote its value at time step $k\in\mb{N} $ by $s_k$. The elements of $s_k\in\mb{R}^{n_\mr{s}}$ are denoted as $s^{[i]}_k$, with $i=1,\dots,n_\mr{s}$. The shift-operator is denoted by $q$, i.e. $q s_k = s_{k+1}$. A data set of $N$ data points taken from the signal $s$, i.e. $\{s_1, s_2, \dots, s_N\}$, is denoted as $\msf{s}=\datasetk{N}{s_k}$, or $\msf{s}$ in short. We denote the column vectorization of $\msf{s}$ with $\col(\msf{s})\in\mb{R}^{n_\mr{s}N}$, i.e. $\col(\msf{s})=\begin{bsmallmatrix} s_1^\top & \cdots & s_N^\top \end{bsmallmatrix}^\top$. The block diagonal matrix with the elements of $\msf{s}$ on the diagonal is denoted as
\begin{equation}
\diag(\msf{s}):= \left[\begin{smallmatrix}
    s_1 &  & 0 \\
     & \sddots &  \\
    0 &  & s_N
\end{smallmatrix}\right]\in\mb{R}^{n_\mr{s}N\times N}.
\end{equation}
The Hankel matrix of row size $L$ associated with the sequence $\msf{s}$ is 
\begin{equation}
\mc{H}_L(\msf{s}):= \left[\begin{smallmatrix}
    s_1 & s_2 & \cdots & s_{N-L+1} \\
    s_2 & s_3 & \cdots & s_{N-L+2} \\
    \svdots & \svdots & \sddots & \svdots \\
    s_L & s_{L+1} & \cdots & s_N
\end{smallmatrix}\right]\in\mb{R}^{n_\mr{s}L\times N-L+1}.\label{eq:hankeldef}
\end{equation}
We denote the predicted values of a variable $s_k$ at time $k + i$ based on the available information at time $k$ as $s_{i|k}$; note that $s_{0|k} = s_k$. %Throughout the paper, the data set $\msf{s}$ will also be referred to as a trajectory of the considered system. 
The Kronecker product is denoted as $\kron$. The block-diagonal Kronecker operator is denoted as $\bkron$, such that for the sequence $\msf{s}$, the notation $(\msf{s}\bkron I_{n_\psi})$ produces the block diagonal matrix
\begin{equation}
(\msf{s}\bkron I_{n_\psi}):= \left[\begin{smallmatrix}
    s_1 \kron I_{n_\psi} &  & 0 \\
     & \sddots &  \\
    0 &  & s_N \kron I_{n_\psi}
\end{smallmatrix}\right]\in\mb{R}^{n_\mr{s}n_\psi N\times Nn_\psi},
\end{equation}
with $I_{n_\psi}$ the identity matrix of size $n_\psi\times n_\psi$. Note that for $n_\psi=1$, $(\msf{s}\bkron I_\psi)=\diag(\msf{s})$.

%===============================================================================
\shortsection{Preliminaries}\label{sec:preliminaries}
Before we present our problem setting and main results, we give a brief re-cap on the existing behavioral results in the LTI case. These results on data-driven analysis and control of LTI systems are the fundament and inspiration for our results presented in this paper. 
For this section, we consider the behavioral language used in \cite{WillemsRapisardaMarkovskyMoor2005}. A system in the behavioral setting is defined as follows \cite[]{PoldermanWillems1997}.
\begin{defn} 
A dynamic system $\Sigma$ is defined as a triple $\Sigma=\left(\mb{T},\mb{W},\mf{B}\right)$ with $\mb{T}$ a subset of $\mb{R}$, called the time axis, $\mb{W}$ a set called the signal space, and $\mf{B}$ a subset of $\mb{W}^\mb{T}$ called the \emph{behavior}, representing the possible solution trajectories of the system ($\mb{W}^\mb{T}$ is standard mathematical notation for the collection of all maps from $\mb{T}$ to $\mb{W}$).
\end{defn}
In this paper we consider finite-dimensional, discrete-time (DT) systems on the time interval $\mb{T}=\mb{N}$ with initial conditions at $k=1$ and $\mb{W}\subseteq\mb{R}^{n_\mr{w}}$. Furthermore, in this section we consider LTI systems, which implies that $\mf{B}$ is linear and shift-invariant, i.e., $q\mf{B}\subseteq \mf{B}$. 
%
% ######################################################################
%
% Regarding the time-invariance of the behavior, we can also use left-compact support of the solution trajectories, s.t. $q\mf{B}=\mf{B}$.
%
% ######################################################################
%
Moreover, we consider systems for which $\mf{B}$ is complete, i.e., closed in the topology of point-wise convergence. Let $\mf{B}_{[1,N]}$ denote the set of signals $w\in\mf{B}$ restricted to the time interval $[1,N]$. We can introduce an IO partitioning of $w_k\in\mb{W}$ in terms of $w_k=[u_k^\top \  y_k^\top ]^\top$, with $u_k\in\mb{U}\subseteq\mb{R}^{n_\mr{u}}$ being free, and $y_k\!\in\mb{Y}\subseteq\mb{R}^{n_\mr{y}}$ being the output (determined by the input, the system and the initial conditions) and $n_\mr{w}=n_\mr{u}+n_\mr{y}$. From \cite{MaRa08}, the \emph{state-space representation} associated with $\mf{B}$ is as
%two important invariant properties of $\mf{B}$,  for which we use the 
\begin{equation}\label{eq:ss_rep}
q x = Ax + Bu; \qquad y = Cx + Du, 
\end{equation}
with $x: \mb{N} \rightarrow \mb{R}^{n_\mathrm{x}}$ the state variable with state dimension $n_\mr{x}>0$ and $A\in \mb{R}^{n_\mathrm{x} \times n_\mathrm{x} }$, $B\in \mb{R}^{n_\mathrm{x} \times n_\mathrm{u} }$, $C\in \mb{R}^{n_\mathrm{y} \times n_\mathrm{x} }$, $D\in \mb{R}^{n_\mathrm{y} \times n_\mathrm{u} }$. The manifest behavior of \eqref{eq:ss_rep} is
\begin{equation*}
\mf{B}'_{A,B,C,D}:=  \{ \coldef{u,y}\in\mf{B} \mid  \exists\, x\in\left(\mb{R}^{n_\mr{x}}\right)^\mb{N}  \text{ s.t. }  \eqref{eq:ss_rep} \text{ holds}\}. 
\end{equation*}
We call \eqref{eq:ss_rep} with order $n_\mr{x}$ a state-space representation of $\mf{B}$ if $\mf{B}'_{A,B,C,D}=\mf{B}$. Next, we introduce two important invariant properties of $\mf{B}$, which are the system order $\mathbf{n}(\mf{B})$, which is the smallest possible order of $\mf{B}'_{A,B,C,D}=\mf{B}$, and the lag $\mathbf{l}(\mf{B})$. The lag is the minimum number of time steps that are required to uniquely determine the initial state using IO data\footnote{Note that $\mathbf{n}(\mf{B})$ and $\mathbf{l}(\mf{B})$ are not necessarily equivalent.}\tss{\!,\!}\footnote{For state-space representations of $\mf{B}$, $n_\ell=\mathbf{l}(\mf{B})$ is the smallest integer for which $\mr{rank}(C\ CA\ \cdots \ CA^{n_\ell-1}) = \mathbf{n}(\mf{B})$.}. Assume that the behaviors of the systems we discuss in this paper are \emph{controllable}, which means that for all $N\in\mb{N}$, $w\in\mf{B}_{[1,N]}$ and $v\in\mf{B}$, there exists always a bridging trajectory $r\in\mf{B}$ and $N'\in\mb{N}$, such that $r_{[1,N]}=w$ and $v_{k-N-N'}=r_k$ for $k>N+N'$, see \cite{WillemsRapisardaMarkovskyMoor2005}. Next, we discuss the notion of persistency of excitation. 
\begin{defn}\label{def:persexcit}
Consider a data sequence $\msf{v}=\{v_k\}_{k=1}^{N_\mr{d}}$, with $v_k\in\mb{R}^{n_\mr{v}}$ and with Hankel matrix $\mc{H}_L(\msf{v})$ as defined in \eqref{eq:hankeldef}. $\msf{v}$ is persistently exciting of order $L$ if $\rank{\mc{H}_L(\msf{v})} = n_\mr{v}L.$
\end{defn}
The condition of persistency of excitation is widely used in system identification and originates from the estimation of FIR filters, where the condition must hold for an input sequence \cite[]{Ljung1999}. Note that Definition \ref{def:persexcit} gives a minimum number for $N_\mr{d}$, namely $N_\mr{d}\!\ge\!(n_\mr{v}+1)L\!-1$. Using the notion of persistence of excitation and the behavioral representation of LTI systems, we formulate Willems' Fundamental Lemma, originating from \cite{WillemsRapisardaMarkovskyMoor2005}. We consider here the version by Berberich and Allg\"ower, adapted for the classical control framework.
\begin{thm}[\cite{BerberichAllgower2020}]\label{thm:thmberberich}
Suppose the sequence $\{u_k, y_k \}_{k=1}^{N_\mr{d}}$ is a trajectory of a controllable LTI system $\Sigma$ with behavior $\mf{B}$ where the input sequence $\msf{u}=\datasetk{N_\mr{d}}{u_k}$ is persistently exciting of order $L+\mathbf{n}(\mf{B})$. Then, $\{\bar{u}_k, \bar{y}_k \}_{k=1}^{L}$ is a trajectory of $\Sigma$, if and only if there exists an $\alpha\in\mb{R}^{N_\mr{d}-L+1}$ such that
\begin{equation}\label{eq:thmberberich}
\begin{bmatrix}
\mc{H}_L(\msf{u}) \\ \mc{H}_L(\msf{y})
\end{bmatrix}\alpha = \begin{bmatrix}
\coldef{\bar{u}} \\ \coldef{\bar{y}}
\end{bmatrix}.
\end{equation}%
%\begin{pf}
%See \cite{BerberichAllgower2020}.
%\end{pf}
\end{thm}
Theorem \ref{thm:thmberberich} means that given $\coldef{u,y}\in\mf{B}_{[1,N_\mr{d}]}$ with $\msf{u}$ persistently exciting of order $L+\mathbf{n}(\mf{B})$, $\coldef{\bar{u},\bar{y}}\in\mf{B}_{[1,L]}$ if and only if there exists $\alpha$ such that \eqref{eq:thmberberich} is satisfied. Hence, all trajectories of length $L$ of a controllable LTI system can be built from linear combinations of time-shifts of a single trajectory from the same LTI system, where the input is persistently exciting. 
%\ttodo{In other words, %The interpretation of this result, is that 
%if the input $u$ is such that all the dynamics of the system are excited (persistently exciting), the time-shifts of the resulting output spans the full output space of the LTI system. Hence, any linear combination of time-shifted versions of the measured input and output will result in a trajectory that admits the system $\Sigma$.} 
In our paper, we aim to extend these results for LPV systems to formulate a data-driven predictive control solution.
%===============================================================================
\shortsection{Data-driven predictor for LPV systems}\label{sec:analysis}
Based on the results in Section \ref{sec:preliminaries}, we now set up the required tools to formulate a data-driven predictive control problem for LPV systems.
\shortsubsection{Considered form of LPV systems}
Consider the DT LPV system with IO representation,
\begin{equation}\label{eq:lpvio}
y_k+{\textstyle\sum_{i=1}^{n_\mr{a}}}\,a_i(p_{k-i})y_{k-i}={\textstyle\sum_{i=1}^{n_\mr{b}}}\,b_i(p_{k-i})u_{k-i},
\end{equation}
where $u_k\in\mb{R}^{n_\mr{u}}$ is the input, $y_k\in\mb{R}^{n_\mr{y}}$ is the output and $p_k\in\mb{P}\subseteq\mb{R}^{n_\mr{p}}$ is the scheduling signal, with $n_\mr{u}$, $n_\mr{y}$, and $n_\mr{p}$ the dimensions of the input, output, and scheduling signals, respectively. $\mb{P}\subseteq\mb{R}^{n_\mr{p}}$ is the scheduling space, which defines the range of the scheduling signal. The behavior of \eqref{eq:lpvio} is defined as
\begin{equation}
\hspace{-0.1mm}\mf{B}_{\mr{LPV}}\!:=\! \{ \coldef{u,p,y}\!\in\! (\mb{R}^{n_\mr{u}}\!\!\times\!\mb{P}\! \times\! \mb{R}^{n_\mr{y}}\!)^\mb{N}\mid \text{s.t.}\, \eqref{eq:lpvio}\, \text{holds}   \}.\hspace{-4mm}
\end{equation}
 The LPV systems in this paper are such that the functions $a_i$ and $b_i$ have the following form\footnote{With this form, the system \eqref{eq:lpvio} can be rewritten into state-space form, where the matrices $\{A,B,C,D\}$ are dependent on the instantaneous value of the scheduling signal \cite[]{AbbasTothWerner2010}.}
\begin{equation}
%a_i(p_{k-i}) = \shortsum{j=0}{n_\mr{p}}a_{i,j}[p_{k-i}]_j, \ \ \ b_i(p_{k-i}) = \sum_{j=0}^{n_\mr{p}}b_{i,j}[p_{k-i}]_j, \label{eq:def_coef}
a_i(p_{k-i}) =  {\textstyle\sum_{j=0}^{n_\mr{p}}}a_i^{[j]}\, p_{k-i}^{[j]}, \ \ \ b_i(p_{k-i}) =  {\textstyle\sum_{j=0}^{n_\mr{p}}}b_i^{[j]}p_{k-i}^{[j]}, \label{eq:def_coef}
%\hspace{-1mm}a_i(p_{k-i}) = {\textstyle\sum_{j=0}^{n_\mr{p}}}a_{i,j}[p_{k-i}]_j, \  b_i(p_{k-i}) = {\textstyle\sum_{j=0}^{n_\mr{p}}}b_{i,j}[p_{k-i}]_j, \hspace{-1mm}\label{eq:def_coef}
\end{equation}
where $p_k^{[0]}=1$ for all $k$. Note that in practice, \eqref{eq:lpvio} often describes a nonlinear system, with e.g., $p_k:=\psi(y_k,u_k)$ \cite[]{To10}. Note that $\mf{B}_{\mr{LPV}}$ is linear in the sense that for any $(u,p,y),(\tilde{u},p,\tilde{y})\in \mf{B}_{\mr{LPV}}$ and $\alpha,\tilde{\alpha}\in\mb{R}$, $(\alpha u+\tilde{\alpha}\tilde{u},p,\alpha y+\tilde{\alpha}\tilde{y})\in \mf{B}_{\mr{LPV}}$. Furthermore, $\mf{B}_{\mr{LPV}}$ is shift-invariant, i.e., $q\mf{B}_{\mr{LPV}}\subseteq \mf{B}_{\mr{LPV}}$. Similar to the LTI case, $\mathbf{n}(\mf{B}_{\mr{LPV}})$ is well-defined and can be computed via a direct minimal state-space realization of \eqref{eq:lpvio}, see  \cite{To10} and \cite{AbbasTothWerner2010}. We assume the following:
\begin{assum}
$\mathbf{n}(\mf{B}_{\mr{LPV}})$ and  $\mathbf{l}(\mf{B}_{\mr{LPV}})$ are known.
\end{assum} 
If \eqref{eq:lpvio} corresponds to a Single-Input-Single-Output (SISO) LPV system (i.e., $n_\mr{u}=n_\mr{y}=1$), and if the left- and right-hand side of \eqref{eq:lpvio} seen as polynomials in the time-shift operator $q$ with coefficient functions dependent on $p$ are co-prime in the Ore algebra defined in \cite[Ch. 3]{To10}, then $\mathbf{n}(\mf{B}_{\mr{LPV}})=\max(n_\mr{a},n_\mr{b})$. In the Multiple-Input-Multiple-Output (MIMO) case, the minimal state order can be determined with the cut \& shift construction of state maps, see \cite[Sec. 4.3]{To10} for more details.
 
\begin{figure}
\centering
\includegraphics[width=\linewidth]{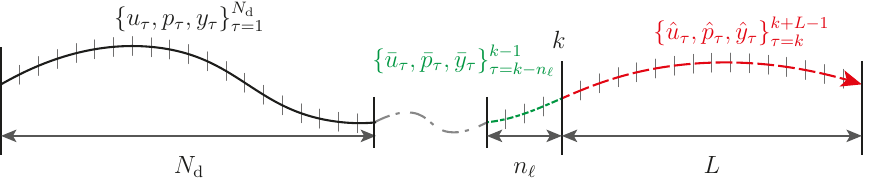}

\ifx\originalspace\undefined \vspace{-3mm} \fi

\caption{Prediction problem with `data-dictionary' ($ \{ u_\tau, p_\tau, y_\tau \}_{\tau=1}^{N_\mr{d}}$, black) of size $N_\mr{d}$, future trajectory ($\{ \hat{u}_\tau, \hat{p}_\tau, \hat{y}_\tau \}_{\tau=k}^{k+L-1}$, red) starting at time step $k$ with prediction horizon $L$ for the LPV system, and the measured $n_\ell$-length initial condition trajectory ($\{ \bar{u}_\tau, \bar{p}_\tau, \bar{y}_\tau \}_{\tau=k-n_\ell}^{k-1}$, green).}
\label{fig:trajectory}
\end{figure}

Consider the situation where we measure the data sequence $\msf{w} = \datasetk{N_\mr{d}}{u_k, p_k,y_k}$ from the LPV system \eqref{eq:lpvio}, for which we assume that the data in $\msf{w}$ is noise-free\footnote{
The extension of our results to include noisy data depends on how the noise influences $(u,p,y)$ and can lead to numerous scenarios. The analysis of these cases is far beyond 
the scope of this paper, where we focus on the formulation of the proposed data-driven approach. } for the sake of simplicity. To formulate a predictive data-driven control scheme, we first extend the theory in Section \ref{sec:preliminaries} to obtain a data-based representation form of \eqref{eq:lpvio}.
% develop a data-based representation form of the system \eqref{eq:lpvio} by the extension of the theory discussed in Section \ref{sec:preliminaries}.
\shortsubsection{Alternative representation form}
We first substitute \eqref{eq:def_coef} in \eqref{eq:lpvio}, which results in the form
\begin{multline}\label{eq:lpviowrittenout}
y_k+{\textstyle\sum_{i=1}^{n_\mr{a}}}a_i^{[0]}y_{k-i} + {\textstyle\sum_{i=1}^{n_\mr{a}}}\bar{a}_i\cdot\left(p_{k-i}\kron y_{k-i}\right)=\\={\textstyle\sum_{i=1}^{n_\mr{b}}}b_i^{[0]}u_{k-i} +{\textstyle\sum_{i=1}^{n_\mr{b}}}\bar{b}_i\cdot\left(p_{k-i}\kron u_{k-i}\right),
%y_k+\sum_{i=1}^{n_\mr{a}}a_{i,0}y_{k-i} + \sum_{i=1}^{n_\mr{a}}\bar{a}_i\cdot\left(p_{k-i}\kron y_{k-i}\right)=\\=\sum_{i=1}^{n_\mr{b}}b_{i,0}u_{k-i} +\sum_{i=1}^{n_\mr{b}}\bar{b}_i\cdot\left(p_{k-i}\kron u_{k-i}\right),
\end{multline}
where $\bar{a}_i:= [ a_i^{[1]} \ \cdots \ a_i^{[n_\mr{p}]} ]$ and $\bar{b}_i:= [b_i^{[1]} \ \hdots \ b_i^{[n_\mr{p}]} ]$. Next, we define new, auxiliary inputs and outputs
\begin{equation}\label{eq:auxIO}
\mc{U}_k := \begin{bmatrix}
u_k \\
p_k \kron u_k
\end{bmatrix}, \quad \mc{Y}_k := \begin{bmatrix}
y_k \\
p_k \kron y_k
\end{bmatrix},
\end{equation}
for the IO system \eqref{eq:lpviowrittenout}, which yields the \emph{implicit} IO form
\begin{equation}\label{eq:lpviorecast}
E\mc{Y}_k+{\textstyle\sum_{i=1}^{n_\mr{a}}}A_i\mc{Y}_{k-i}={\textstyle\sum_{i=1}^{n_\mr{b}}}B_i\mc{U}_{k-i}, 
\end{equation}
with $E= [I \ 0]$, $A_i := [a_i^{[0]} \ \bar{a}_{i}]$ and $B_i := [b_i^{[0]} \ \bar{b}_{i}]$. Rewriting the LPV system representation in this way is commonly utilized in LPV IO identification, e.g.,  \cite{LaGiToGa10}, and LPV subspace identification, e.g.,
\cite{WingerdenVerhaegen2009,CoxToth2021}.
%Rewriting an LPV system as such has earlier been proposed for identification of LPV systems in \cite{LaGiToGa10}, and a similar approach in LPV subspace identification, see e.g. \cite{WingerdenVerhaegen2009,CoxToth2021}.
 Furthermore, note that this augmentation does not influence the order of the system \cite[]{WollnackEtAl2017}. 
\begin{note}\label{note:ttp}
Let the data set $\{p_1 \kron s_1, p_2 \kron s_2, \dots , p_{N_\mr{d}} \kron s_{N_\mr{d}}\}$ of $N_\mr{d}$ points be denoted as $\msf{s}^\mt{p}=\datasetk{N_\mr{d}}{p_k \otimes s_k}$ or $\msf{s}^\mt{p}$ in short. We can similarly denote the column vectorization of $\msf{s}^\mt{p}$ by $\col(\msf{s}^\mt{p})\in\mb{R}^{n_\mr{p}n_\mr{s}N_\mr{d}}$. Note that $\col(\msf{s}^\mt{p})=(\msf{p}\bkron I_{n_\mr{s}})\col(\msf{s})$.
\end{note}
From Note \ref{note:ttp}, the data-sequences satisfying \eqref{eq:lpviorecast} are $\msf{u}$, $\msf{u}^\mt{p}$, $\msf{y}$, and $\msf{y}^\mt{p}$. The sequences of the auxiliary inputs and outputs \eqref{eq:auxIO} are denoted $\msf{U}=\datasetk{N_\mr{d}}{\mc{U}_k}$ and $\msf{Y}=\datasetk{N_\mr{d}}{\mc{Y}_k}$.
\shortsubsection{Application of Willems' Fundamental Lemma}\label{sec:applicationWillemsLemma}
Note that the implicit IO form \eqref{eq:lpviorecast} is LTI without the constraint \eqref{eq:auxIO}, where the latter is only a restriction of the associated LTI behavior $\mf{B}$ of \eqref{eq:lpviorecast}, i.e.,
\begin{equation}
 \mf{B}_{\mr{LPV}} = \{ (\mc{Y}, \mc{U} ) \in \mf{B}   \mid \exists p\in\mb{P}^\mb{N} \text{ s.t. }  \eqref{eq:auxIO} \text{ holds}  \}
\end{equation}
This gives the core idea to apply the established theory by Willems et al. to obtain a data-driven LPV system representation. Suppose we obtain the set of data $\datasetk{N_\mr{d}}{u_k, p_k, y_k}$ from the LPV system, with the auxiliary input sequence $\{\mc{U}_k\}_{k=1}^{N_\mr{d}}$ being persistently exciting of order $n_\mr{x}+L$, i.e., \mbox{$\rank{\mc{H}_{n_\mr{x}+L}(\msf{U})}=(n_\mr{p}+1)n_\mr{u}(n_\mr{x}+L)$}. We then can apply Willems' Fundamental Lemma on \eqref{eq:lpviorecast}. Let $\msf{U}$ be persistently exciting of order $n_\mr{x}+L$, then $\datasetk{L}{\bar{u}_k,\bar{p}_k, \bar{y}_k }$ is a trajectory of the original LPV system if there exists a $g\in\mb{R}^{N_\mr{d}-L+1}$, such that
\begin{equation}\label{eq:willemslemmaext}
\newcommand{\vphsp}{\vphantom{\msf{u}^{\mt{p}^k}}\!}
\begin{bmatrix}
\mc{H}_L\left( \msf{U} \right) \\
\mc{H}_L\left( \msf{Y} \right) 
\end{bmatrix}g = \begin{bmatrix}
\coldef{\bar{\msf{U}}} \\
\coldef{\bar{\msf{Y}}}
\end{bmatrix}\!  \Leftrightarrow \! \begin{bmatrix}
\mc{H}_L\left( \msf{u} \right) 		\vphsp \\
\mc{H}_L\left( \msf{u}^\mt{p} \right) 	\vphsp \\
\mc{H}_L\left( \msf{y} \right) 		\vphsp \\
\mc{H}_L\left( \msf{y}^\mt{p} \right) 	\vphsp
\end{bmatrix} \! g = \! \begin{bmatrix}
\coldef{\bar{\msf{u}}} 				\vphsp \\
\coldef{\bar{\msf{u}}^{\bar{\mt{p}}}} 	\vphsp \\
\coldef{\bar{\msf{y}}} 				\vphsp \\
\coldef{\bar{\msf{y}}^{\bar{\mt{p}}}} 	\vphsp 
\end{bmatrix}\! ,
\end{equation} 
and condition \eqref{eq:auxIO} is satisfied, which means that
\[ \coldef{\bar{\msf{y}}^{\bar{\mt{p}}}}  = (\bar{\msf{p}}\bkron I_{n_\mr{y}})\col(\bar{\msf{y}}), \ \text{and} \ \ \coldef{\bar{\msf{u}}^{\bar{\mt{p}}}}  = (\bar{\msf{p}}\bkron I_{n_\mr{u}})\col(\bar{\msf{u}}). \]
From \eqref{eq:willemslemmaext}, it follows that
\begin{equation}
\mc{H}_{L}( \msf{y} )g=\col(\bar{\msf{y}}), \text{ and }\ \mc{H}_{L}( \msf{y}^\mt{p} )g=\col(\bar{\msf{y}}^{\bar{\mt{p}}}),
\end{equation}
which combined by the above relations gives
\begin{equation*}
\mc{H}_{L}( \msf{y}^\mt{p} )g=\col(\bar{\msf{y}}^{\bar{\mt{p}}})=(\bar{\msf{p}}\bkron I_{n_\mr{y}})\col(\bar{\msf{y}})=(\bar{\msf{p}}\bkron I_{n_\mr{y}})\mc{H}_{L}( \msf{y} )g, 
\end{equation*}
i.e., $\left[\mc{H}_{L}\left( \msf{y}^\mt{p} \right) -(\bar{\msf{p}}\bkron I_{n_\mr{y}})\mc{H}_{L}\left( \msf{y} \right)\right]g=0$, together with a similar relationship for $\bar{\msf{u}}$. Let $\bar{\mc{P}}^{n_\mr{u}}=(\bar{\msf{p}}\bkron I_{n_\mr{u}})$ and $\bar{\mc{P}}^{n_\mr{y}}=(\bar{\msf{p}}\bkron I_{n_\mr{y}})$. Combining these relations gives the data-driven representation of \eqref{eq:lpvio}, which is the first main contribution: 
\begin{equation}\label{eq:LPV_pred}
\begin{bmatrix}
\mc{H}_L\left( \msf{u} \right) 		\\
\mc{H}_{L}\left( \msf{u}^\mt{p} \right) -\bar{\mc{P}}^{n_\mr{u}}\mc{H}_{L}\left( \msf{u} \right) \\
\mc{H}_L\left( \msf{y} \right) 		\\
\mc{H}_{L}\left( \msf{y}^\mt{p} \right) -\bar{\mc{P}}^{n_\mr{y}}\mc{H}_{L}\left( \msf{y} \right)
\end{bmatrix}g = \begin{bmatrix}
\coldef{\bar{\msf{u}}} 				\\
0	 \\
\coldef{\bar{\msf{y}}} 				 \\
0	 
\end{bmatrix},
\end{equation}
where the matrix in the left-hand side of the equality is of dimension $L(n_\mr{u}+n_\mr{y})(1+n_\mr{p})\times (N_\mr{d}-L+1)$, while $g\in\mb{R}^{N_\mr{d}-L+1}$ and the vector in the right-hand side of the equality is of dimension $L(n_\mr{u}+n_\mr{y})(1+n_\mr{p})\times 1$. 
\shortsubsection{Data-driven LPV predictor}
As a next step, we use  \eqref{eq:willemslemmaext} to formulate a data-driven LPV predictor, which is suitable for LPV system simulation, and the design of a predictive controller. Consider the prediction problem depicted in Fig. \ref{fig:trajectory}. To ensure the smooth continuation of the future trajectory at time step $k$, we need to take the initial conditions of the future trajectory into account. For a certain future input and scheduling trajectory, there are still an infinite number of possible output trajectories. By `attaching' an initial trajectory to the predicted one, we can determine a unique output trajectory, which continues smoothly after time step $k$. Note that the length of this initial trajectory must be larger than $\mathbf{l}(\mf{B}_{\mr{LPV}})$ \cite[]{MaRa08}.
Let $n_\ell$ and $n_\mr{x}$ be equal to the lag and the order of the LPV system, respectively. Now, given a sequence of data $\{u_\tau, p_\tau, y_\tau\}_{\tau=1}^{N_\mr{d}}$, a trajectory $\{\bar{u}_\tau, \bar{p}_\tau, \bar{y}_\tau\}_{\tau=k-n_\ell}^{k-1}$ of length $n_\ell$, and a data sequence of length $L$ of future input and scheduling trajectories $\{\hat{u}_\tau, \hat{p}_\tau\}_{\tau=k}^{k+L-1}$. The goal is to determine the corresponding output sequence $\{\hat{y}_\tau\}_{\tau=k}^{k+L-1}$. The sequences used at time $k$ are denoted with subscript $k$, i.e., $\hat{\msf{y}}_k := \{\hat{y}_\tau\}_{\tau=k}^{k+L-1}$. The problem in \eqref{eq:willemslemmaext} can now be split up, such that $L$ in \eqref{eq:willemslemmaext} is substituted with the length of the initial trajectory and the predicted trajectory, which yields Hankel matrices such as $\mc{H}_{L+n_\ell}( \msf{s})$. We can split-up such a Hankel matrix as 
$$\mc{H}_{n_\ell+L}(\msf{s})=\begin{bmatrix}\mc{H}^\top_{n_\ell+L,n_\ell}(\msf{s}) & \mc{H}^\top_{n_\ell+L,L}(\msf{s}) \end{bmatrix}^\top,$$
such that $\mc{H}_{n_\ell+L,n_\ell}(\msf{s})$ contains the first $n_\ell\, n_\mr{s}$ rows of $\mc{H}_{L+n_\ell}(\msf{s})$, and $\mc{H}_{n_\ell+L,L}(\msf{s})$ contains the last $L \,n_\mr{s}$ rows of $\mc{H}_{L+n_\ell}(\msf{s})$. Splitting up \eqref{eq:willemslemmaext} yields
\begin{equation}\label{eq:willemslemmasplit}
\begin{bmatrix}
\mc{H}_{n_\ell+L,n_\ell}\left( \msf{U} \right) \\
\mc{H}_{n_\ell+L,n_\ell}\left( \msf{Y} \right) \\
\mc{H}_{n_\ell+L,L}\left( \msf{U} \right) \\
\mc{H}_{n_\ell+L,L}\left( \msf{Y} \right)
\end{bmatrix}g = \begin{bmatrix}
\col(\bar{\msf{U}}_k) \\
\col(\bar{\msf{Y}}_k) \\
\col(\hat{\msf{U}}_k) \\
\col(\hat{\msf{Y}}_k)
\end{bmatrix}\begin{array}{l}
\left. \vphantom{\begin{matrix}
\col(\bar{\msf{U}}_k) \\
\col(\bar{\msf{Y}}_k)
\end{matrix}}\right\} \text{Initial trajectory} \\
\left. \vphantom{\begin{matrix}
\col(\hat{\msf{U}}_k) \\
\col(\hat{\msf{Y}}_k)
\end{matrix}}\right\} \text{Future trajectory.}
\end{array}\hspace{-2mm}
\end{equation} 
Note that each `line' in \eqref{eq:willemslemmasplit}, e.g. $\mc{H}_{n_\ell+L,n_\ell}\left( \msf{U} \right) g = \col(\bar{\msf{U}}_k)$, results in two equalities, i.e., $\mc{H}_{n_\ell+L,n_\ell}\left( \msf{u} \right) g = \col(\bar{\msf{u}}_k)$ and $\mc{H}_{n_\ell+L,n_\ell}\left( \msf{u}^\mt{p} \right) g = \col(\bar{\msf{u}}_k^{\bar{\mt{p}}})$. Application of \eqref{eq:LPV_pred} allows to compute the future sequence $\hat{\msf{y}}_k=\{\hat{y}_\tau\}_{\tau=k}^{k+L-1}$ in terms of the following steps:
%
%
%
%Consider the resulting equalities of the last `line' in \eqref{eq:willemslemmasplit}, i.e.
%\begin{equation}
%\mc{H}_{L}( \msf{y} )g=\col(\hat{\msf{y}}), \text{ and }\ \mc{H}_{L}( \msf{y}^\mt{p} )g=\col(\hat{\msf{y}}^{\hat{\mt{p}}}).
%\end{equation}
%Next, we combine the two equalities using $\col(\hat{\msf{y}}^{\hat{\mt{p}}})=\diag(\hat{\msf{p}})\col(\hat{\msf{y}})$, from Note \ref{note:ttp}, as follows
%\begin{equation*}
%\mc{H}_{L}( \msf{y}^\mt{p} )g=\col(\hat{\msf{y}}^{\hat{\mt{p}}})=\diag(\hat{\msf{p}})\col(\hat{\msf{y}})=\diag(\hat{\msf{p}})\mc{H}_{L}( \msf{y} )g, 
%\end{equation*}
%i.e. $\left[\mc{H}_{L}\left( \msf{y}^\mt{p} \right) -\diag(\hat{\msf{p}})\mc{H}_{L}\left( \msf{y} \right)\right]g=0$. This trick eliminates the term $\hat{\msf{y}}$ from the computational problem. This yields that we can compute the future sequence $\hat{\msf{y}}$ using the following steps:
\begin{enumerate}
\item Determine a solution $g$ of
\begin{equation}\label{eq:hossamstrick1}
\hspace{-5mm}\begin{bmatrix} 
\mc{H}_{n_\ell+L,n_\ell}\left( \msf{u} \right) \\
\mc{H}_{n_\ell+L,n_\ell}\left( \msf{u}^\mt{p} \right) -\bar{\mc{P}}^{n_\mr{u}}_k\mc{H}_{n_\ell+L,n_\ell}\left( \msf{u} \right)\\
\mc{H}_{n_\ell+L,n_\ell}\left( \msf{y} \right) \\
\mc{H}_{n_\ell+L,n_\ell}\left( \msf{y}^\mt{p} \right) -\bar{\mc{P}}^{n_\mr{y}}_k\mc{H}_{n_\ell+L,n_\ell}\left( \msf{y} \right) \\
\mc{H}_{n_\ell+L,L}\left( \msf{u} \right) \\
\mc{H}_{n_\ell+L,L}\left( \msf{u}^\mt{p} \right) -\hat{\mc{P}}^{n_\mr{u}}_k\mc{H}_{n_\ell+L,L}\left( \msf{u} \right) \\
\mc{H}_{n_\ell+L,L}\left( \msf{y}^\mt{p} \right) -\hat{\mc{P}}^{n_\mr{y}}_k\mc{H}_{n_\ell+L,L}\left( \msf{y} \right)
\end{bmatrix} g =
\begin{bmatrix}
\col(\bar{\msf{u}}_k) \\
0 \\
\col(\bar{\msf{y}}_k) \\
0 \\
\col(\hat{\msf{u}}_k) \\
0 \\
0
\end{bmatrix},\hspace{-5mm}
\end{equation}
where $\bar{\mc{P}}^{n_\mr{u}}_k:= (\bar{\msf{p}}_k\bkron I_{n_\mr{u}})$, $\bar{\mc{P}}^{n_\mr{y}}_k:= (\bar{\msf{p}}_k\bkron I_{n_\mr{y}})$, $\hat{\mc{P}}^{n_\mr{u}}_k:= (\hat{\msf{p}}_k\bkron I_{n_\mr{u}})$ and $\hat{\mc{P}}^{n_\mr{y}}_k:= (\hat{\msf{p}}_k\bkron I_{n_\mr{y}})$.
\item Calculate $\col(\hat{\msf{y}}_k)=\begin{bmatrix} \hat{y}_k^\top & \cdots & \hat{y}_{k+L-1}^\top \end{bmatrix}^\top$ from
\begin{equation}\label{eq:hossamstrick2}
\col({\hat{\msf{y}}_k}) = \mc{H}_{n_\ell+L,L}\left( \msf{y} \right)g.
\end{equation}
\end{enumerate}
These two steps are the key in order to perform LPV data-driven simulation and for formulating the predictor in the proposed predictive control scheme, which is explained in the next section, where the length $L$ of the future trajectory is the prediction horizon (denoted by $N_\mr{p}$).

%===============================================================================
\shortsection{LPV data-driven predictive control}\label{sec:dpc}
LPV model-based predictive control is a practical control approach for controlling nonlinear and time-varying systems which can be formulated into LPV representations. However, due to its model-based nature, it requires an LPV predictor describing the dynamics of the system which is obtained via physical modeling and LPV embedding methods or LPV system identification. To avoid the step of developing an LPV model for the system, which is usually a tedious process, we propose in this section a data-driven predictive control   algorithm for controlling LPV systems without information about the system other than its order and given short trajectories of its input, output and scheduling signals, such that the persistence of excitation condition as in Definition \ref{def:persexcit} is satisfied.

It is worth to mention that the application of predictive control for LPV systems requires the knowledge of the future scheduling signal over the prediction horizon of the controller, which is often a priori not available. The simplest way to handle this is to freeze the scheduling signal over the prediction horizon at its current value, leading to so-called \emph{gain-scheduled} MPC. To avoid such a crude approximation, LPV-MPC solutions often apply a robustness based characterization of the future scheduling variable \cite[]{Zappa2003}, iterative simulation based synthesis of the future scheduling trajectory \cite[]{CISNEROS201711601}, scheduling prediction \cite[]{Morato19}, or a tube based construction \cite[]{Hanema2021}.

With the ingredients of Section~\ref{sec:analysis}, we are ready to setup the proposed LPV-DPC\footnote{Stability and feasibility guarantees of the proposed predictive control scheme are not discussed in this paper, as we concentrate here only on the formulation.} using the data-driven LPV predictor. We consider tracking of a reference signal $r:\mb{N} \rightarrow \mb{R}^{n_\mr{y}}$ under a quadratic loss function. Consider a so-called data dictionary $\{u_l, p_l, y_l\}_{l=1}^{N_\mr{d}}$, denoted $(\msf{u}, \msf{p}, \msf{y})$, where $\msf{u}$ is persistently exciting of $N_\mr{p}+n_\mr{x}$, and the recent observation past $\{u_{\tau},p_{\tau},y_{\tau}\}_{\tau=k-n_\ell}^{k-1}$ is denoted $(\bar{\msf{u}}_k, \bar{\msf{p}}_k, \bar{\msf{y}}_k)$. 

Let the prediction horizon $N_\mr{p}$ be equal to the control horizon $N_\mr{c}$ for the sake of simplicity. Let $\mb{U}$, $\mb{Y}$ denote the input and output constraint sets. Assume that the future trajectory of the scheduling $\hat{\msf{p}}_k=\{p_{i|k}\}_{i=0}^{N_\mr{p}}$ is known till the end of the prediction horizon $N_\mr{p}$. Construct $\hat{\mc{P}}^{n_\mr{u}}_k$, $\hat{\mc{P}}^{n_\mr{y}}_k$, $\bar{\mc{P}}^{n_\mr{u}}_k$ and $\bar{\mc{P}}^{n_\mr{y}}_k$ as in Step (1).
For a quadratic tracking error loss and input penalty characterized in terms of $Q,R \succeq 0$, the quadratic DPC optimization problem can be formulated as follows:
\begin{subequations}\label{eq:lpvdpc}
\begin{align}
\hspace{-2mm}\min_{g} & \ \textstyle{\sum_{i=0}^{N_\mr{p}-1}}
 (\hat{y}_{i|k}-r_{i|k})^\top Q (\hat{y}_{i|k}-r_{i|k}) 
 + \hat{u}_{i|k}^\top R\, \hat{u}_{i|k} \hspace{-2mm} \\
 \text{ s.t.} &
  \; \col(\hat{\msf{u}}_k) =  \begin{bmatrix} 
 \hat{u}_{0|k} & \cdots & \hat{u}_{N_\mr{p}-1 |k} \end{bmatrix}^\top\!\! \in \mb{U}^{N_\mr{p}-1} \\
&
\mc{H}_{n_\ell+N_\mr{p},N_\mr{p}}
\left( \msf{y} \right)g  =  \begin{bmatrix} 
 \hat{y}_{0|k} & \cdots & \hat{y}_{N_\mr{p}-1 |k} \end{bmatrix}^\top\!\! \in \mb{Y}^{N_\mr{p}-1} \hspace{-2mm}\\
& \hspace{-3mm}
\begin{bmatrix} 
\mc{H}_{n_\ell+N_\mr{p},n_\ell}\left( \msf{u} \right) \hphantom{\qquad\quad\quad }\\
\!\!\mc{H}_{n_\ell+N_\mr{p},n_\ell}\!\!\left( \msf{u}^\mt{p} \right) \!\unaryminus\bar{\mc{P}}^{n_\mr{u}}_k\mc{H}_{n_\ell+N_\mr{p},n_\ell}\!\!\left( \msf{u} \right)\!\\
\mc{H}_{n_\ell+N_\mr{p},n_\ell}\left( \msf{y} \right) \hphantom{\qquad\quad\quad }\\
\!\!\mc{H}_{n_\ell+N_\mr{p},n_\ell}\!\!\left( \msf{y}^\mt{p} \right) \!\unaryminus\bar{\mc{P}}^{n_\mr{y}}_k\mc{H}_{n_\ell+N_\mr{p},n_\ell}\!\!\left( \msf{y} \right) \!\\
\mc{H}_{n_\ell+N_\mr{p},N_\mr{p}}\left( \msf{u} \right) \hphantom{\qquad\quad\quad }\\
\!\mc{H}_{n_\ell+N_\mr{p},N_\mr{p}}\!\!\left( \msf{u}^\mt{p} \right) \!\unaryminus\hat{\mc{P}}^{n_\mr{u}}_k\mc{H}_{n_\ell+N_\mr{p},N_\mr{p}}\!\!\left( \msf{u} \right)\! \\
\!\mc{H}_{n_\ell+N_\mr{p},N_\mr{p}}\!\!\left( \msf{y}^\mt{p} \right) \!\unaryminus\hat{\mc{P}}^{n_\mr{y}}_k\mc{H}_{n_\ell+N_\mr{p},N_\mr{p}}\!\!\left( \msf{y} \right)\!
\end{bmatrix} \! g \!=\!\!
\begin{bmatrix}
\!\col(\bar{\msf{u}}_k)\! \\
0 \\
\!\col(\bar{\msf{y}}_k)\! \\
0 \\
\!\col(\hat{\msf{u}}_k)\! \\
0 \\
0
\end{bmatrix}\!\!,\label{eq:lpvdpc_cond3}\!\!
\end{align}
\end{subequations}
where in \eqref{eq:lpvdpc_cond3}, $g\in \mb{R}^{N_\mr{d}-n_\ell-N_\mr{p}+1}$ and the right-hand side is in $\mb{R}^{(n_\ell+N_\mr{p})(n_\mr{u}+n_\mr{y})(1+n_\mr{p})-N_\mr{p}n_\mr{y}}$.
In this DPC scheme, the data dictionary $(\msf{u},\msf{y},\msf{p})$ is given in advance, and is used to construct the left-hand side of \eqref{eq:lpvdpc_cond3}. Note that $(\msf{u},\msf{p})$ satisfies the persistence of excitation condition as in Section \ref{sec:applicationWillemsLemma}, and that the data dictionary is \emph{fixed} throughout operation, i.e. the persistence of excitation condition remains guaranteed during operation. The observation past $\bar{\msf{u}}_k, \bar{\msf{y}}_k, \bar{\msf{p}}_k$, is \emph{updated} at every time instant $k$ from recent input/output/scheduling measurements from the system, that is, $\bar{\msf{u}}_{k-1}, \bar{\msf{y}}_{k-1}, \bar{\msf{p}}_{k-1}$ is \emph{replaced} with $\bar{\msf{u}}_k, \bar{\msf{y}}_k, \bar{\msf{p}}_k$. The sequences $\hat{\msf{u}}_k$ and $\hat{\msf{y}}_k$ are dependent decision variables, which are connected to the decision variable $g$ and the data matrices $\mc{H}_{n_\ell+N_\mr{p},N_\mr{p}}(\cdot)$. Important hyper-parameters of the scheme are $N_\mr{p}$ and $n_\ell$. For the former, the design considerations are similar as for standard MPC solutions. The latter must be chosen as large as possible to satisfy both $n_\ell \geq \mathbf{l}(\mf{B}_{\mr{LPV}})$ and $N_\mr{d}\ge((n_\mr{p}+1)n_\mr{u}+1)(\mathbf{n}(\mf{B}_{\mr{LPV}})+N_\mr{p})-1$. However, this increases the number of constraints in \eqref{eq:lpvdpc_cond3} and hence the computational load.

In the introduced scheme the trajectory of the scheduling $\hat{\msf{p}}_k=\{p_{i|k}\}_{i=0}^{N_\mr{p}}$ in the prediction horizon is assumed to be known. As it was previously mentioned, this is a central problem in LPV predictive control and all
%\cout{available strategies in terms of gain-scheduling:  $p_{i|k}=p_k$ for all $i\in\{1,\ldots,N_\mr{p}\}$, robust formulation of \eqref{eq:lpvdpc} with rate of variation bounds on $p_k$ or the iterative scheme discussed in Cisneros and Werner (2017)}
% \cite{CISNEROS201711601} } 
the aforementioned existing methodologies can be directly utilized in our LPV-DPC setting.

%Moreover, $\hat{\msf{p}}$, which is used to construct $\hat{\msf{u}}^{\hat{\mt{p}}}$ and $\hat{\msf{y}}^{\hat{\mt{p}}}$ 
%could  be given in advance, if its future information is available; otherwise, it can be frozen over the prediction horizon\footnote{It is common  in the literature to consider the scheduling signals over $N_\mr{p}$ as an uncertain parameter, then robust predictive control is invoked which usually is  computationally costly. Several approaches have been proposed to reduce such  computation burden, see \cite{MoNoSe20} for more details.}. 
%In practice, in case of \ttodo{quasi}-LPV case, obtaining these values of $\hat{\msf{p}}$ in advance can be handled in an iterative procedure, e.g., the approach of \cite{CiVoWe16}. 

%
% of the above data-driven LPV system representation as the predictor in a predictive control scheme for reference tracking. Stability and feasibility will not be discussed here. As an initial step of this approach we assume that the future scheduling parameter trajectory and the reference $r$ are given in advance. First we discuss the optimization problem.

%{\color{red} Maybe we add the Algorithm}
%===============================================================================
\shortsection{Numerical example}\label{sec:example}
We give two examples to show the effectiveness of the proposed approach.
The simulations here have been carried out using \textsc{Matlab} R2020a on 64-bit Windows 10, with Gurobi 8.0 as the
Quadratic Program (QP) solver.
%Next, we illustrate the performance of the proposed LPV data-predictive control method using a numerical example taken from the literature \cite{LaGiToGa10} and we compare the  performance of the LPVDPC with that of the LPVMPC for the same numerical example. 

\shortsubsection{Example 1: Academic example}
In this example we illustrate the performance of the proposed LPV data-predictive control method when it is applied on an LPV system where the scheduling variable depends only on  exogenous signals, i.e., independent of the system dynamics. We consider a numerical example taken from  \cite{LaGiToGa10}  on which we compare the  performance of the LPV-DPC with an LPV-MPC design. 

\begin{figure}[t!]
\centering
\includegraphics[width=\linewidth]{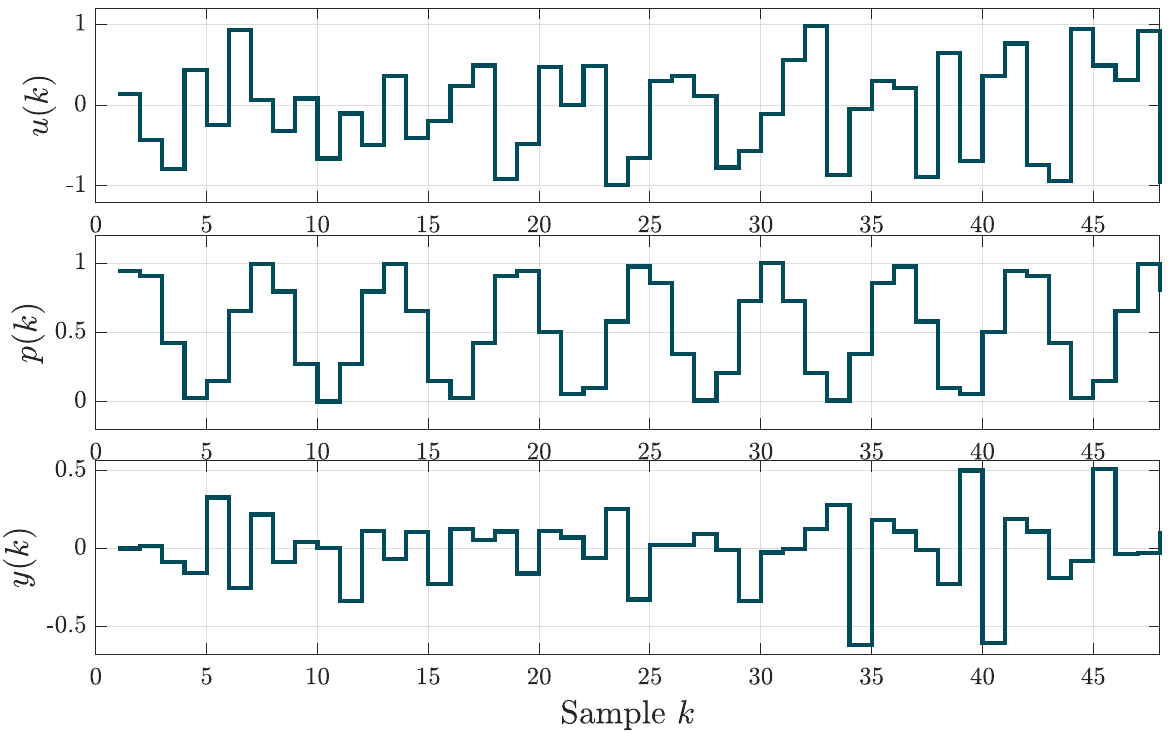}

\ifx\originalspace\undefined \vspace{-3mm} \fi

\caption{Data dictionary of Example 1. Trajectories of $u, p, y$ used to construct the Hankel matrices for \eqref{eq:lpvdpc_cond3}.}
\label{fig:traj_example1}
\end{figure}

Consider a SISO LPV system represented by \eqref{eq:lpvio} with $n_\mr{a}=2,n_\mr{b}=2$ and
%\begin{align*}
%a_1(p_k) &= 1 - 0.5p_k -0.1 p_k^2 \\
%a_2(p_k) &= 0.5 - 0.7p_k -0.1 p_k^2 \\
%b_1(p_k) &= 0.5 - 0.4p_k +0.01 p_k^2 \\
%b_2(p_k) &= 0.2 - 0.3p_k -0.2 p_k^2.
%\end{align*}
\begin{align*}
a_1(p_k) &\!=\! 1 \unaryminus 0.5p_k \unaryminus0.1 p_k^2 & 
a_2(p_k) &\!=\! 0.5 \unaryminus 0.7p_k \unaryminus0.1 p_k^2 \\
b_1(p_k) &\!=\! 0.5 \unaryminus 0.4p_k \!+\!0.01 p_k^2  & 
b_2(p_k) &\!=\! 0.2 \unaryminus 0.3p_k \unaryminus0.2 p_k^2.
\end{align*}
For this system, $\mathbf{n}(\mf{B}_{\mr{LPV}})=n_\mr{a}=n_\mr{x}=2$ and  $\mathbf{l}(\mf{B}_{\mr{LPV}})=2$. In this example, we choose $n_\ell=2$ and $N_\mr{p}=5$, we treat $p_k$ and $p_k^2$ as the scheduling variables, i.e., $n_\mr{p}=2$, and we set $N_\mr{d}=48$. The scheduling space is defined as $\mb{P}:= [0,\,1]\times[0,\,1]$. 
%\ttodo{Do we have a single signal $p_k$, which is also squared in the model? Beacuse then this is a NL system. If these are the first and second elements of a 2-dim scheduling signal, then it is an LPV system... we need to make this clear here, now it kind of floats around} {\color{blue} Below it is written that we consider them as two scheduling parameters}\\
The data dictionary $\datasetk{N_\mr{d}}{u_k, p_k, y_k}$, shown in Fig.~\ref{fig:traj_example1}, is generated from the system using a uniformly distributed input with $p_k=0.5\sin(0.35\pi k)+0.5$. This data is used to construct the data sets $\msf{u}, \msf{u}^\mt{p}, \msf{y}, \msf{y}^\mt{p}$ as shown in \eqref{eq:auxIO} and \eqref{eq:willemslemmaext}, which are used offline to construct the Hankel matrices for the equality constraint \eqref{eq:lpvdpc_cond3}. These matrices should satisfy the persistence of excitation condition and are fixed during online operation. The past observations are updated at every time step using the past data of $u,y,p$ at instants $k-1,k-2$. The resulting constraint \eqref{eq:lpvdpc_cond3} is a function of the decision variable $g$ and the future values of $\col(\hat{\msf{p}}_k)$ over the prediction horizon. We assume here that the trajectory of $p_k$ is available in advance.
%; otherwise, they can be predicted using tools from e.g. machine learning.
 The constraint sets are $\mathbb{U}:=[\unaryminus5,\; 5]$ and $\mathbb{Y}:=[\unaryminus1,\; 1]$. For the LPV-MPC, the exact model is used with the same choice of $N_\mr{p}=5$ and given the future values of the scheduling variables. The cost matrices  $Q=10$ and  $R=0.001$ are used for both the LPV-DPC and the LPV-MPC. We have carried out the simulation for both LPV-MPC and LPV-DPC algorithms controlling the system to track a given reference trajectory. According to the receding horizon principle, in both cases, the first instance $\hat{u}_k$ of the computed optimal input sequence at every time instant $k$ is applied to the system.
\begin{figure}[t!]
\centering
\includegraphics[width=\linewidth]{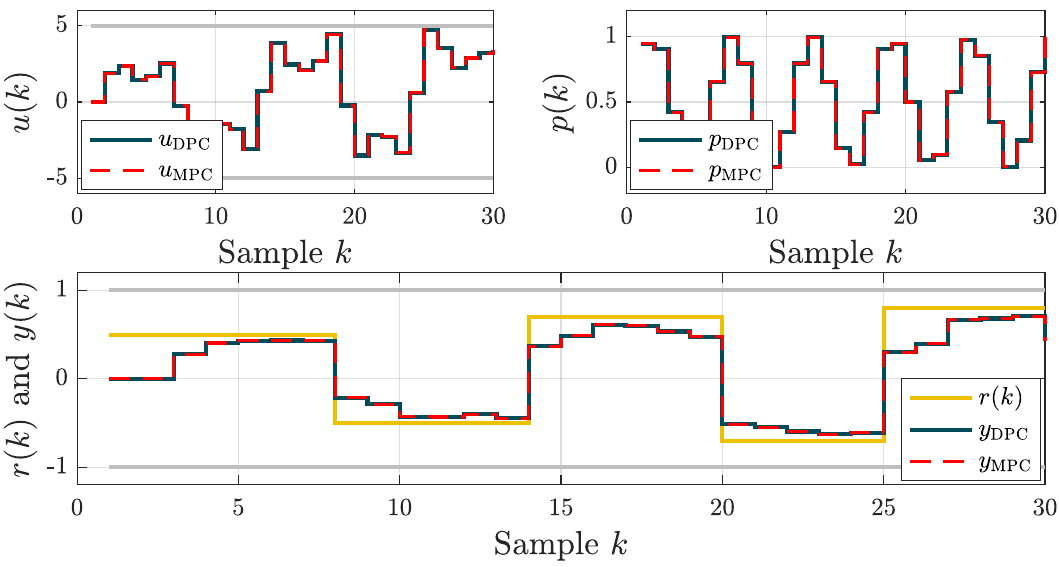}

\ifx\originalspace\undefined \vspace{-3mm} \fi

\caption{Results of Example 1. Control input (\textsc{nw}), scheduling variable (\textsc{ne}) and reference tracking (\textsc{s}).}
\label{fig:res_example1}
\end{figure}

The reference tracking results together with the control input and scheduling trajectory used during both simulations are shown in  Fig.~\ref{fig:res_example1}. It can be observed that both the LPV-DPC and the LPV-MPC provide indistinguishable results.
This demonstrates that our proposed DPC is an exact alternative to MPC, provided that the data is noise-free.
%{\color{red} do we need to include the data-driven simulation results?}
%\setlength\figH{3cm} 
%\setlength\figW{4cm}
\shortsubsection{Example 2: DC Motor with Unbalanced Disc}
To illustrate the effectiveness of the LPV-DPC when it is used for controlling nonlinear systems  and to compare its performance with  LPV-MPC  we consider the problem of controlling  a simulator of a DC motor connected to an unbalanced disc.  The behavior of the system corresponds to a rotational pendulum; this example has been taken from \cite{BoCoTo21}. The system is nonlinear and it can be represented by the following differential equation:
\begin{equation}\label{eq:NLunbaldisc}
\ddot{\theta}(t)=-\frac{mgl}{J}\sin(\theta(t))-\frac{1}{\tau}\dot{\theta}(t)+\frac{K_\mr{m}}{\tau}u(t),
\end{equation}
 where $\theta$ is angular position of the disc,   $u$ the input voltage to the system, which is its control input, and
 $m,g,l,J,\tau,K_\mr{m}$ are the physical parameters of the system, as shown in Table \ref{tab:unbaldisc_param}.
%\begin{table}
%\centering
%\caption{Parameters of the unbalanced disc.}
%\label{tab:unbaldisc_param}
%\begin{tabular}{l|c|r|r}
%Parameter & \multicolumn{1}{l|}{Symbol} & \multicolumn{1}{l|}{Value} & \multicolumn{1}{l}{Unit} \\ \hline
%Gravitational acceleration & $g$ & $9.8$&  $[\mr{m}\cdot\mr{s}^{\unaryminus2}]$    \\
%Disc inertia & $J$ & $2.2\cdot10^{\unaryminus4}$ & $[\mr{Nm}^2]$ \\
%Motor constant & $K_\mr{m}$ & $15.3145$ & $[\unaryminus]$ \\
%Distance mass to center & $l$ & $0.42$ & $[\mr{mm}]$ \\
%Lumped mass & $m$ & $0.07$ & $[\mr{kg}]$ \\
%Back-EMF constant & $\tau$ & $0.5971$ & $[\unaryminus]$     \\
%Sampling time & $T_\mr{s}$ & $75\cdot10^{\unaryminus3}$ &  $[\mr{s}]$
%\end{tabular}
%\end{table}
\begin{table}[!ht]
\centering
\caption{Parameters of the unbalanced disc.}
\label{tab:unbaldisc_param}{\small
\begin{tabular}{l||c|c|c|c|c|c|c}
Parameter \!\!\!\!\! & $g$ & $J$ &\!\!\! $K_\mr{m}$\!\!\!& $l$ & $m$ & $\tau$ & $T_\mr{s}$ \\ \hline
Value & $9.8$&\!\!\!$2.2\!\cdot\!10^{\!\unaryminus\!4}$\!\!\! & \!\!\!$15.3$ \!\!\!&\!\!\!$0.42$\!\!\! &\!\!\!$0.07$\!\!\! &$0.6$ & $75$ \\
Unit & \!\!\!$[\mr{m}\!\cdot\!\mr{s}^{\!\unaryminus\!2}]$\!\!\!& $[\mr{Nm}^2]$& $[\unaryminus]$ &\!\!\!$[\mr{mm}]$ \!\!\!& $[\mr{kg}]$&   $[\unaryminus]$&  $[\mr{ms}]$ 
\end{tabular}}
\end{table}

We aim here at controlling the system using the proposed LPV-DPC and compare its performance with the  LPV-MPC. The output $y(t)=\theta(t)$ of the closed-loop system should follow a predefined reference trajectory without violating its input and output limits, which are given by the constraint sets $\mb{U} :=[\unaryminus \tfrac{1}{4},\,\tfrac{1}{4}]$, $\mb{Y} :=[\unaryminus 1,\, 1]$,
% \begin{equation}\label{eq:limits_NLunbaldisc}
% \mb{U} :=[\unaryminus \tfrac{1}{4},\,\tfrac{1}{4}] ,\quad \mb{Y} :=[\unaryminus 1,\, 1]
% \end{equation}
 respectively.

The LPV embedding of the nonlinear dynamics of the system can be carried out by introducing
\begin{equation}\label{eq:sch_NLunbaldisc}
p(t)={\rm sinc}(\theta(t))=\frac{\sin(\theta(t))}{\theta(t)}\in\mb{P}
\end{equation}
as  the scheduling variable, which implies $\mb{P}:=[\unaryminus 0.22,\, 1]$. %such that \eqref{eq:limits_NLunbaldisc} is and $n_\mr{x}=2$, respectively, satisfied

\begin{figure}[t!]
\centering
\includegraphics[width=\linewidth]{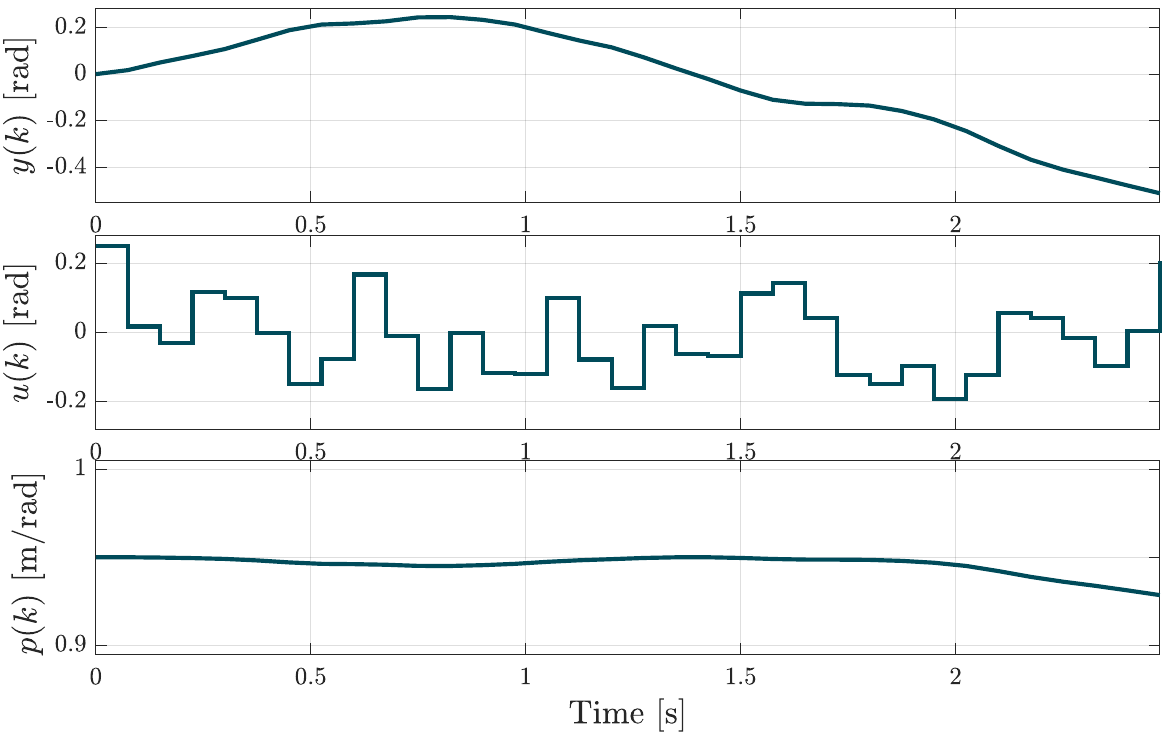}

\ifx\originalspace\undefined \vspace{-3mm} \fi

\caption{Data dictionary of Example 2: Trajectories of $y, u, p$ used to construct Hankel matrices in Example 2.}
\label{fig:traj_unbaldisc}
\end{figure}

 The first step to solve the optimization problem \eqref{eq:lpvdpc} of the LPV-DPC  is to generate offline the trajectories $\datasetk{N_\mr{d}}{u_k, p_k, y_k}$ from the system.  Therefore, we have simulated \eqref{eq:NLunbaldisc} for a random-phase multi-sine input with $N_\mr{d}=34$ and a sampling time of $75$ ms. The generated trajectories are shown in Fig.~\ref{fig:traj_unbaldisc}. 
 Next, these trajectories are used to construct  the column vectors
$\col(\msf{u}), \col(\msf{y}), \col(\msf{p})$. For this system, $\mathbf{n}(\mf{B}_{\mr{LPV}})=2$ and $n_\ell =\mathbf{l}(\mf{B}_{\mr{LPV}})=2$, and $N_\mr{p}=5$ are chosen to compute the corresponding Hankel matrix as shown in the left hand side of \eqref{eq:lpvdpc_cond3}  such that the corresponding persistently excitation   condition is satisfied. 
The column vectors on the right-hand side of \eqref{eq:lpvdpc_cond3} are constructed online, similar to the previous example. The decision variable in this case is $g\in\mathbb{R}^{28}$. Since the system here is represented as an LPV system where the scheduling variable depends on endogenous signals of the system, the future scheduling trajectory $\col(\hat{\msf{p}}_k)\in\mathbb{R}^5$ over the prediction horizon  is a function of $\hat{\msf{y}}$ according to \eqref{eq:sch_NLunbaldisc}, which is not available in advance. For practical implementation, we consider gain-scheduling, i.e., freezing the scheduling variable at its current measured value $p_k$ over the prediction horizon, which often leads to efficient controllers based on solving a simple QP problem at every sample. To avoid complexity, we consider here this simple approach. % Then, all required trajectories to construct the equality constraint  \eqref{eq:lpvdpc_cond3} become  available; for $\breve{\msf{u}},{\breve{\msf{u}}^{\breve{\mt{p}}}}, \breve{\msf{y}}, {\breve{\msf{y}}^{\breve{\mt{p}}}}$, they can be updated at every time $k$  using recent data of $u,y,p$. 
%{\color{red} We should mention that this way is very different from just using LTI data at every $k$, namely, local LTI model as 
%here we have global Hankel matrices and the intitalization vectors $\breve{\msf{u}},{\breve{\msf{u}}^{\breve{\mt{p}}}}, \breve{\msf{y}}, {\breve{\msf{y}}^{\breve{\mt{p}}}}$ have full information about recent $p$. We should also mention that we frame work we consider here is general with dynamic dependence scheduling signal according to the formulation using (4).}

The cost function is chosen with $Q=0.1$ and $R=0.05$. For the purpose of comparison, we also control the system using LPV-MPC based on a DT state-space LPV representation of  \eqref{eq:NLunbaldisc} computed with a $4^\text{th}$ Runge-Kutta method and sampling time of $75$ ms. Similar to the LPV-DPC, for the LPV-MPC, we considered that $p$ is frozen over the prediction horizon. Next, both controllers, the  proposed  LPV-DPC and the LPV-MPC,  have been implemented on the nonlinear continuous-time simulator of \eqref{eq:NLunbaldisc} with zero-order-hold and with the objective of reference tracking from an initial condition $\theta(0)=\unaryminus0.9$ [rad] and $\dot\theta(0)=0$ [rad/s]. 
\begin{figure}[h!]
\centering
\includegraphics[width=\linewidth]{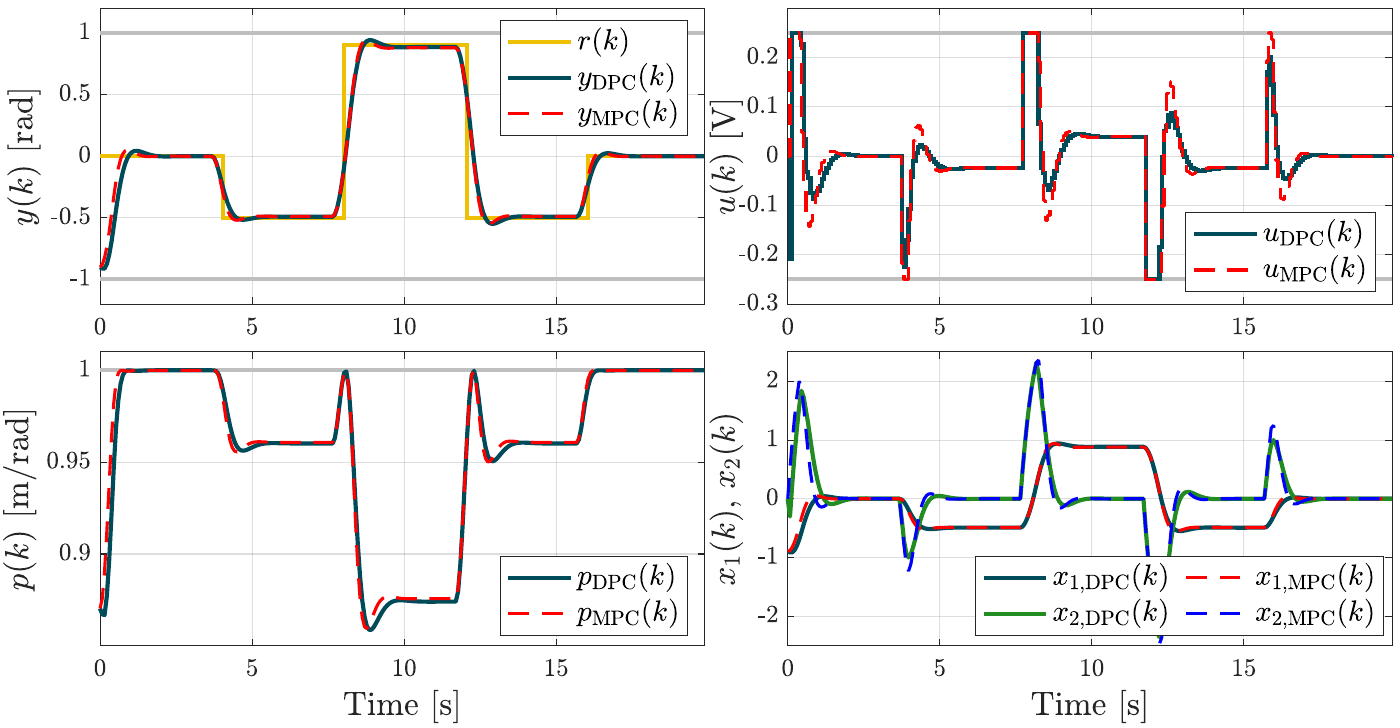}

\ifx\originalspace\undefined \vspace{-3mm} \fi

\caption{Results of Example 2: reference tracking, control input, scheduling variable and state evolution of the system; reference (yellow), LPV-DPC (blue) and LPV-MPC (red-dashed).}
\label{fig:res_unbaldisc}
\end{figure}
The resulted output, input, scheduling and state trajectories of the system with both controllers are shown in Figs.~\ref{fig:res_unbaldisc}. These results show that both controllers provide  similar performance without violating the IO constraints. This simulation demonstrates that the proposed model-free LPV-DPC can control the nonlinear system efficiently and exactly as the LPV-MPC without any information other than a single trajectory of 34 samples and the knowledge that the system is of order two, while the LPV-MPC requires a full system model. 
%{\color{red} Matlab R2020a on 64-bit Windows 10, with Gurobi 8.0 is the QP solver using Yalmip}
 %\setlength\figH{3cm} 
%\setlength\figW{4cm}

%===============================================================================
\shortsection{Conclusions and outlook}\label{sec:conclusion}
In this paper, a novel data-driven predictive control scheme for reference tracking control of unknown LPV systems has been introduced. Formulation of this control approach is made possible by our novel result to derive an LPV data-driven predictor based  on the behavioral system theory. The predictor only requires reasonably short recorded trajectories of the system input, output and scheduling signals, to represent the complete dynamic behavior of the system if a persistence of excitation condition is satisfied. 
The introduced predictive control related optimization problem is a single QP subject to an equality constraint representing the predictor in addition to input and output constraints.
Numerical simulations has shown that the proposed LPV-DPC is comparable to LPV-MPC for time-varying and nonlinear systems. As a future work we will focus on establishing stability and feasibility guarantees of the proposed LPV-DPC method and analyze its behavior under noisy data.
%===============================================================================

%% There are a number of predefined theorem-like environments in
%% ifacconf.cls:
%%
%% \begin{thm} ... \end{thm}            % Theorem
%% \begin{lem} ... \end{lem}            % Lemma
%% \begin{claim} ... \end{claim}        % Claim
%% \begin{conj} ... \end{conj}          % Conjecture
%% \begin{cor} ... \end{cor}            % Corollary
%% \begin{fact} ... \end{fact}          % Fact
%% \begin{hypo} ... \end{hypo}          % Hypothesis
%% \begin{prop} ... \end{prop}          % Proposition
%% \begin{crit} ... \end{crit}          % Criterion

\bibliography{references}

\begin{thebibliography}{29}
\providecommand{\natexlab}[1]{#1}
\providecommand{\url}[1]{\texttt{#1}}
\providecommand{\urlprefix}{URL }
\expandafter\ifx\csname urlstyle\endcsname\relax
  \providecommand{\doi}[1]{doi:\discretionary{}{}{}#1}\else
  \providecommand{\doi}{doi:\discretionary{}{}{}\begingroup
  \urlstyle{rm}\Url}\fi

\bibitem[{Abbas et~al.(2010)Abbas, T\'oth, and Werner}]{AbbasTothWerner2010}
Abbas, H.S., T\'oth, R., and Werner, H. (2010).
\newblock {State-Space Realization of LPV Input--Output Models: Practical
  Methods for The User}.
\newblock In \emph{Proc. of the 2010 American Control Conference}.

\bibitem[{Berberich and Allg\"ower(2020)}]{BerberichAllgower2020}
Berberich, J. and Allg\"ower, F. (2020).
\newblock A trajectory-based framework for data-driven system analysis and
  control.
\newblock In \emph{Proc. of the 2020 European Control Conference}.

\bibitem[{Chisci et~al.(2003)Chisci, Falugi, and Zappa}]{Zappa2003}
Chisci, L., Falugi, P., and Zappa, G. (2003).
\newblock {Gain-scheduling MPC of nonlinear systems}.
\newblock \emph{International Journal of Robust and Nonlinear Control}.

\bibitem[{Cisneros and Werner(2017)}]{CISNEROS201711601}
Cisneros, P.G. and Werner, H. (2017).
\newblock {Fast Nonlinear MPC for Reference Tracking Subject to Nonlinear
  Constraints via Quasi-LPV Representations}.
\newblock In \emph{Proc. of the 20{th} IFAC World Congress}.

\bibitem[{Coulson et~al.(2019)Coulson, Lygeros, and
  D{\"o}rfler}]{CoulsonLygerosDorfler2019}
Coulson, J., Lygeros, J., and D{\"o}rfler, F. (2019).
\newblock {Data-enabled predictive control: In the shallows of the DeePC}.
\newblock In \emph{Proc. of the 2019 European Control Conference}.

\bibitem[{Cox and T{\'o}th(2021)}]{CoxToth2021}
Cox, P. and T{\'o}th, R. (2021).
\newblock Linear parameter-varying subspace identification: A unified
  framework.
\newblock \emph{Automatica}, 123, 109296.

\bibitem[{De~Persis and Tesi(2019{\natexlab{a}})}]{dePersisTesi2019journal}
De~Persis, C. and Tesi, P. (2019{\natexlab{a}}).
\newblock {Formulas for Data-Driven Control: Stabilization, Optimality, and
  Robustness}.
\newblock \emph{IEEE Transactions on Automatic Control}.

\bibitem[{De~Persis and Tesi(2019{\natexlab{b}})}]{DePersisTesi2019}
De~Persis, C. and Tesi, P. (2019{\natexlab{b}}).
\newblock {On Persistency of Excitation and Formulas for Data-Driven Control}.
\newblock In \emph{Proc. of the 58{th} Conference on Decision and Control}.

\bibitem[{den Boef et~al.(2021)den Boef, Cox, and T\'oth}]{BoCoTo21}
den Boef, P., Cox, P.B., and T\'oth, R. (2021).
\newblock {LPVcore: MATLAB Toolbox for LPV Modelling, Identification and
  Control}.
\newblock \emph{Accepted to the 19{th} IFAC Symposium on System
  Identification}.

\bibitem[{Formentin et~al.(2013)Formentin, Piga, T{\'o}th, and
  Savaresi}]{formentin2013direct}
Formentin, S., Piga, D., T{\'o}th, R., and Savaresi, S.M. (2013).
\newblock Direct data-driven control of linear parameter-varying systems.
\newblock In \emph{Proc. of the 52{nd} IEEE Conference on Decision and
  Control}.

\bibitem[{Formentin et~al.(2016)Formentin, Piga, T{\'o}th, and
  Savaresi}]{Formentin2016direct}
Formentin, S., Piga, D., T{\'o}th, R., and Savaresi, S.M. (2016).
\newblock {Direct learning of LPV controllers from data}.
\newblock \emph{Automatica}.

\bibitem[{Hanema et~al.(2021)Hanema, T{\'o}th, and Lazar}]{Hanema2021}
Hanema, J., T{\'o}th, R., and Lazar, M. (2021).
\newblock Stabilizing non-linear model predictive control using linear
  parameter-varying embeddings and tubes.
\newblock \emph{IET Control Theory \& Applications}.

\bibitem[{Hewing et~al.(2020)Hewing, Wabersich, Menner, and
  Zeilinger}]{HeWaMeZe20}
Hewing, L., Wabersich, K.P., Menner, M., and Zeilinger, M.N. (2020).
\newblock Learning-based model predictive control: Toward safe learning in
  control.
\newblock \emph{Annual Review of Control, Robotics, and Autonomous Systems}.

\bibitem[{Hoffmann and Werner(2015)}]{HoWe15}
Hoffmann, C. and Werner, H. (2015).
\newblock A survey of linear parameter-varying control applications validated
  by experiments or high-fidelity simulations.
\newblock \emph{IEEE Transactions on Control Systems Technology}.

\bibitem[{Hou and Wang(2013)}]{HouWang2013}
Hou, Z.S. and Wang, Z. (2013).
\newblock From model-based control to data-driven control: Survey,
  classification and perspective.
\newblock \emph{Information Sciences}.

\bibitem[{Laurain et~al.(2010)Laurain, Gilson, T\'oth, and
  Garnier}]{LaGiToGa10}
Laurain, V., Gilson, M., T\'oth, R., and Garnier, H. (2010).
\newblock {Refined Instrumental Variable Methods for Identification of LPV
  Box--Jenkins Models}.
\newblock \emph{Automatica}.

\bibitem[{Ljung(1999)}]{Ljung1999}
Ljung, L. (1999).
\newblock \emph{{System Identification: Theory for the User}}, volume~26.
\newblock Prentice Hall PTR, Upper Saddle River, NJ, 2 edition.

\bibitem[{Maciejowski(2002)}]{Ma02j}
Maciejowski, J. (2002).
\newblock \emph{Predictive Control with Constraints}.
\newblock Pearson Education Limited.

\bibitem[{Markovsky and Rapisarda(2008)}]{MaRa08}
Markovsky, I. and Rapisarda, P. (2008).
\newblock {Data-Driven Simulation and Control}.
\newblock \emph{International Journal of Control}.

\bibitem[{Morato et~al.(2020)Morato, Normey-Rico, and Sename}]{MoNoSe20}
Morato, M.M., Normey-Rico, J.E., and Sename, O. (2020).
\newblock Model predictive control design for linear parameter varying systems:
  A survey.
\newblock \emph{Annual Reviews in Control}.

\bibitem[{Morato et~al.(2019)Morato, Normey-Rico, and Sename}]{Morato19}
Morato, M.M., Normey-Rico, J.E., and Sename, O. (2019).
\newblock {Novel qLPV MPC Design with Least-Squares Scheduling Prediction}.
\newblock In \emph{Proc. of the 3rd IFAC Workshop on Linear Parameter-Varying
  Systems}. Eindhoven, The Netherlands.

\bibitem[{Oomen et~al.(2013)Oomen, van Herpen, Quist, van~de Wal, Bosgra, and
  Steinbuch}]{oomen2013connecting}
Oomen, T., van Herpen, R., Quist, S., van~de Wal, M., Bosgra, O., and
  Steinbuch, M. (2013).
\newblock Connecting system identification and robust control for
  next-generation motion control of a wafer stage.
\newblock \emph{IEEE Transactions on Control Systems Technology}.

\bibitem[{Polderman and Willems(1997)}]{PoldermanWillems1997}
Polderman, J.W. and Willems, J.C. (1997).
\newblock \emph{{Introduction to Mathematical Systems Theory: A Behavioral
  Approach}}, volume~26.
\newblock Springer.

\bibitem[{Romer et~al.(2019)Romer, Berberich, K\"ohler, and
  Allg\"ower}]{RomerBerberichKohlerAllgower2019}
Romer, A., Berberich, J., K\"ohler, J., and Allg\"ower, F. (2019).
\newblock {One-shot verification of dissipativity properties from input-output
  data}.
\newblock \emph{IEEE Control Systems Letters}.

\bibitem[{T\'oth(2010)}]{To10}
T\'oth, R. (2010).
\newblock \emph{Modeling and Identification of Linear Parameter-Varying
  Systems}.
\newblock Springer-Verlag.
\newblock 1\tss{st} ed.

\bibitem[{Van~Wingerden and Verhaegen(2009)}]{WingerdenVerhaegen2009}
Van~Wingerden, J.W. and Verhaegen, M. (2009).
\newblock {Subspace identification of Bilinear and LPV systems for open-and
  closed-loop data}.
\newblock \emph{Automatica}.

\bibitem[{Willems et~al.(2005)Willems, Rapisarda, Markovsky, and
  De~Moor}]{WillemsRapisardaMarkovskyMoor2005}
Willems, J.C., Rapisarda, P., Markovsky, I., and De~Moor, B.L.M. (2005).
\newblock A note on persistency of excitation.
\newblock \emph{Systems \& Control Letters}.

\bibitem[{Wollnack et~al.(2017)Wollnack, Abbas, T{\'o}th, and
  Werner}]{WollnackEtAl2017}
Wollnack, S., Abbas, H.S., T{\'o}th, R., and Werner, H. (2017).
\newblock {Fixed-structure LPV-IO controllers: An implicit representation based
  approach}.
\newblock \emph{Automatica}.

\bibitem[{Zhou et~al.(1996)Zhou, Doyle, and Glover}]{zhou1996robust}
Zhou, K., Doyle, J.C., and Glover, K. (1996).
\newblock \emph{{Robust and Optimal Control}}, volume~40.
\newblock Prentice hall New Jersey.

\end{thebibliography}

\end{document}